\documentclass[conference]{IEEEtran}
\IEEEoverridecommandlockouts
\usepackage[noend]{algpseudocode}
\usepackage{amsmath,amssymb,amsfonts}
\usepackage[labelformat=simple]{subcaption}

\usepackage[hidelinks]{hyperref}
\usepackage[utf8]{inputenc}
\usepackage{dblfloatfix}
\usepackage{tcolorbox}
\usepackage{algorithm}
\usepackage{listings}
\usepackage{graphicx}
\usepackage{booktabs}
\usepackage{textcomp}
\usepackage{graphicx}
\usepackage{cleveref}
\usepackage{multirow}
\usepackage{textcomp}
\usepackage{balance}
\usepackage{comment}
\usepackage{siunitx}
\usepackage{diagbox}
\usepackage{xcolor}
\usepackage{pifont}
\usepackage{bm}
\usepackage{diagbox}

\usepackage[
backend=biber
,style=ieee
,minbibnames=1
,maxbibnames=99
,maxcitenames=1
,mincitenames=1
,eprint=false
,doi=false
,isbn=false
,url=false
]{biblatex}
\bibliography{ref} 
\AtBeginBibliography{\footnotesize}

\def\BibTeX{{\rm B\kern-.05em{\sc i\kern-.025em b}\kern-.08em
    T\kern-.1667em\lower.7ex\hbox{E}\kern-.125emX}}
\begin{document}

\title{Towards Automatic Generation of Amplified Regression Test Oracles}

\author{
\IEEEauthorblockN{
	Alejandra Duque-Torres\IEEEauthorrefmark{2}, Claus Klammer\IEEEauthorrefmark{3}, Dietmar Pfahl\IEEEauthorrefmark{2},
    Stefan Fischer\IEEEauthorrefmark{3}, and Rudolf Ramler\IEEEauthorrefmark{3}}
			
	\IEEEauthorblockA{%
	\IEEEauthorrefmark{2}\textit{Institute of Computer Science  }, \textit{University of Tartu}, Tartu, Estonia \\
	E-mail: \{duquet, dietmar.pfahl\}@ut.ee}	
	\IEEEauthorblockA{
	\IEEEauthorrefmark{3}\textit{Software Competence Center Hagenberg (SCCH) GmbH}, Hagenberg, Austria \\
	E-mail:  \{claus.klammer, stefan.fischer, rudolf.ramler\}@scch.at}
}

\maketitle

\begin{abstract}
Regression testing is crucial in ensuring that pure code refactoring does not adversely affect existing software functionality, but it can be expensive, accounting for half the cost of software maintenance. Automated test case generation reduces effort but may generate weak test suites. Test amplification is a promising solution that enhances tests by generating additional or improving existing ones, increasing test coverage, but it faces the test oracle problem. To address this, we propose a test oracle derivation approach that uses object state data produced during System Under Test (SUT) test execution to amplify regression test oracles. The approach monitors the object state during test execution and compares it to the previous version to detect any changes in relation to the SUT's intended behaviour. Our preliminary evaluation shows that the proposed approach can enhance the detection of behaviour changes substantially, providing initial evidence of its effectiveness.
\end{abstract}


\section{Introduction}
\label{sec:Intro}
Software testing is an essential activity in the software development process as it helps to ensure the correct functioning of the final software \cite{hilton_large-scale_2018}. However, testing has been recognised as a time-consuming and, thus, expensive task due to the size and complexity of modern software systems \cite{QuASoQ}. To tackle these issues, test automation has emerged as an efficient way to improve test efficiency and effectiveness. 
Despite its benefits, test automation faces the Test Oracle problem, which refers to the difficulty in determining the correct output of a software system for a given input \cite{QuASoQ}. Although extensive research has been conducted in this area, the problem remains largely unresolved and presents a significant difficulty for automated testing, particularly in the context of regression testing \cite{patel2019partial}. 


Regression testing is a crucial type of software testing, which involves `retesting' the system or its components to verify that the existing functionality is not adversely affected by any changes, and still meets the specified requirements \cite{10.1002/stv.430}. A system is said to regress when new code is added, or a modification to the existing code introduces faults or affects other parts of the system unexpectedly. Therefore, it is crucial to retest not only the changed code but also the possible affected code due to the change. However, regression testing is an expensive activity, typically accounting for half of the total cost of software maintenance \cite{desikan2006software}. Automated test case generation has emerged as a promising solution to address this issue \cite{patel2019partial}. However, while automated testing can reduce the effort required for regression testing, it also produces rather weak tests that may not fully capture system behaviour changes due to the test oracle problem, which can limit the effectiveness of automated testing in detecting faults and verifying correct system behaviour.

To address the limitations of automated testing in detecting faults and to increase the test coverage in regression testing, test amplification or test augmentation is a promising solution \cite{DANGLOT2019110398}. Test amplification generates additional test cases or enhances existing ones to create stronger tests that better capture changes in system behaviour by increasing the test coverage. However, it also faces the test oracle problem. Therefore, in test amplification, accurate test oracles are essential to ensure that changes made to the system do not introduce unintended behaviour \cite{10.1145/3550355.3552451}.  Motivated by the above, we developed an approach to derive test oracles based on information contained in object state data produced during the test execution of the system under test (SUT). We call such oracles \textit{amplified regression test oracles}. 


In our context, the term \textit{object state data} refers to the data stored within the object after a public method has been called, plus all return values of public methods \cite{QuASoQ}. There are three main types of methods that can interact with object state data. The first type of methods are known as ``setters". These methods are used to modify the object's state by changing the values of the instance variables. 
The second type of methods are known as ``getters" and do not change the object's state data. Instead, they retrieve information about the current state and return a value based on that information. 
Lastly, there are methods that do both, modify and inspect the object's state data. These methods can change the object's state by modifying its state data and return information about the new state data. We call this type of methods ``getter-setters".

The key idea of our approach is to always monitor the methods that return information about the object state data, even if such a method is not executed by a test. 
To achieve this, we use a Test Adapter, which is a program that captures and stores the state data of objects from those methods during test execution. This allows us to observe and analyse the object state data changes throughout the test execution. Since our approach is meant for regression testing, we capture the state data of multiple SUT versions. We then compare the state data obtained during the current version of the SUT to the state data obtained during the previous version. This allows us to determine if any relations in the state data have changed.
If the relations remain unchanged, we can assume that the SUT is behaving correctly and the changes made in the new version of the SUT have not introduced any unintended behaviour. 

When utilising our approach to derive amplified regression test oracles, there are two crucial points that need to be taken into consideration. Firstly, the Test Adapter is designed to capture and store the objects state data by internally executing all methods that may return information about the state data only. Secondly, the Test Adapter should be triggered during the test execution after every test sequence. It is important to note that the Test Adapter cannot work without at least one test. Without a test execution, there is no state information to capture and compare. It is important to emphasise that our approach is not intended to replace the existing test suite. Instead, it is meant to amplify the test suite by comparing the state information captured during the test execution of two different versions of the SUT. 

In this paper, we aim to answer the following research questions:

\textbf{RQ$_{1}$: {How can object state data be used to derive test oracles during test execution?}}  This research question explores the use of object state data to derive test oracles during test execution. It aims to investigate the feasibility of monitoring the object state as the SUT runs and capturing and storing the object state data at every point in time during test execution when a public method is called. 

\textbf{RQ$_{2}$: What is the effectiveness of the amplified regression test oracles in detecting unintended behaviour in the SUT?} This research question aims to investigate the extent of improvement that can be achieved using amplified regression test oracles. Specifically, it seeks to understand the potential benefits of using such oracles to detect unintended behaviour in the SUT.

As a proof-of-concept, we decided to utilise a custom implementation of the Collection Classes in the Java programming language as the SUT in our experiments. The reason for selecting these particular classes is that they exhibit a well-understood state behaviour, making it easier to manipulate and analyse during the experimentation phase. 
By leveraging this simple yet effective approach, we hope to demonstrate the potential benefits of incorporating object state monitoring into existing test suites.

\section{Methodology}
\label{sec:methodology}

This section aims to provide a detailed overview of our approach. In \Cref{subsec:Approach}, we first comprehensively describe each step of our approach. In \Cref{subsec: Example}, we provide three real-world scenarios where our approach can be applied. We also provide a hypothetical situation to demonstrate how our approach can effectively solve a particular problem. To further evaluate the effectiveness of our approach, in \Cref{subsubsec:SUT}, we discuss our evaluation methodology and describe the SUT in which our approach was evaluated.

\subsection{Approach}
\label{subsec:Approach}

Overall, our approach for amplifying regression test oracles comprises two phases. \textit{Phase I} involves generating the Test Adapter and obtaining the initial version of the object's state data. The regression test oracles, represented by the state data of the SUT, are the output of this phase. \textit{Phase II}, amplifies the initial test suite. This is achieved by comparing the previous version of the state data with the new version. The output of this phase can be seen as a fault report for new versions of the SUT.

\subsubsection{\textbf{Phase I}}
\label{subsubsec:PhaseI}
This phase focuses on preparing the initial test suite and gathering the initial object state data. This process can be split into three main steps, which are detailed below:

\textbf{\textit{Step 1.1 - Produce initial test:}} In this step, an initial test suite needs to be generated. By a ``initial test suite", we refer to a small set of tests that is essential for our approach to work. Without at least one test, our approach cannot be executed. It is important to highlight that our approach is designed to add additional checks on top of the initial test suite using the state data. Any automation tool, such as Randoop\footnote{https://randoop.github.io/randoop/} or EvoSuite\footnote{https://www.evosuite.org/} can be used to generate the initial test suite. 

\textbf{\textit{Step 1.2 - State data acquisition:}} in this step, the focus is on monitoring the states of the SUT during the execution of the test suite and saving the state data. To achieve this, we propose using a Test Adapter, which serves as an interface between the SUT and the test suite, allowing for monitoring the SUT's object state during the test execution. The Test Adapter captures information from the SUT's object state by internally executing all methods that may return information about the state data only. It is essential to carefully select the methods that will provide information about the state data to ensure they represent the SUT's state and can provide useful information for later analysis.


\textbf{\textit{Step 1.3 - Test execution:}} During this test suite execution, the Test Adapter is triggered, and it starts monitoring the state of the SUT during the execution of a test sequence at each interaction with the SUT via public method call. The monitoring process captures object state data of the SUT, reflecting the behaviour of the SUT during the test execution, and saves it a CSV file.

\subsubsection{\textbf{Phase II}} This phase involves amplifying the test suite and detecting faults in new versions of the SUT. It consists of two main steps, which are detailed below: 

\textit{\textbf{Step 2.1 - re-execution of test suite:}} In this step, the focus is on re-executing the initial test suite on the new version of the SUT. The main goal is obtaining a new state data version using the Test Adapter from Step 1.2. It is important to note that this step is performed in the context of regression testing, which aims to identify any new defects or regressions that may have been introduced due to changes made to the SUT. During the re-execution of the test suite, the Test Adapter continuously monitors and records the SUT's state data. The output of this step is the new version of the state data.

\textbf{\textit{Step 2.2 - Amplified regression test oracle}}: In this step, the object state data obtained in Step 1.2 is compared with the new version of the state data obtained in Step 2.1. This comparison allows us to identify differences or changes between the two versions of state data. These differences may indicate the presence of faults or regressions in the new version of the SUT. This information can help developers quickly pinpoint the source of the problem and resolve it. 

\subsection{Motivational toy example}
\label{subsec: Example}
Overall, our approach of deriving regression test oracles based on object state data can be helpful in scenarios where it is important to ensure that changes made to the code do not introduce unintended behaviour and where a comprehensive test suite may not be possible. Below, we highlight three of those scenarios:

\begin{itemize}
    \item \textbf{Limited test suite:} In some cases, creating a comprehensive test suite that covers all possible scenarios may not be possible. This could be due to limited resources, time constraints, or the system's complexity. By using our approach to derive test oracles based on object state data, we can ensure that the limited test suite is still effective in catching any unintended behaviour that may occur due to changes made to the code.
    \item \textbf{Refactoring:}  When refactoring, it is crucial to avoid unintended behaviour by ensuring that changes made to the code do not create any issues. A weak existing test suite can make it challenging to spot unintended behaviour. Monitoring the object state data during test execution can help ensure that changes made to the code do not introduce unintended behaviour. If the object state data remains the same as the previous version during test execution, there is a higher likelihood that no unintended behaviour has been introduced.
    \item \textbf{Integration testing:} When integrating different components or systems, ensuring that their interactions do not introduce unintended behaviour is important. Monitoring object state data during integration testing is an effective approach to ensure no unintended behavior is introduced. If the object state data during integration testing remains unchanged from the previous version, it can be concluded with high probability that there's no unintended behavior.
\end{itemize}
\begin{algorithm}
\caption{ArrayCalculator class}
\label{alg:toyExample}
\begin{algorithmic}[1]
\State \textbf{class} ArrayCalculator:
\State \hspace{0.2cm} \textbf{def \_init\_($self$, data):} 
\State \hspace{0.4cm} $self$.data = data \Comment{{\small \textit{Method that initialises the object}}}
\State \hspace{0.2cm} \textbf{def append$(self$, new\_data):}
\State \Comment{{\small \textit{Method that modifies the OS}}}
\State \hspace{0.2cm} \hspace{0.2cm} \textbf{for} $element$ \textbf{in} new\_data: 
\State \hspace{0.2cm} \hspace{0.2cm} \hspace{0.2cm} $self$.data $+= [element]$
\State \hspace{0.2cm} \textbf{def} \textbf{reverse($self$):}  
\State \hspace{0.2cm} \hspace{0.1cm} $s = 0$ \Comment{{\small \textit{Method that modifies the OS}}}
\State \hspace{0.2cm} \hspace{0.1cm} $e = $$self$$.$get\_size() $- 1$
\State \hspace{0.2cm} \hspace{0.1cm}\textbf{while} s $<$ e: \textcolor{red}{\#~mutant: $``>"$}
\State \hspace{0.2cm} \hspace{0.4cm}{\small $self$.data$[s]$, $self$.data$[e] =$ $self$.data$[e]$, $self$.data$[s]$}
\State \hspace{0.2cm} \hspace{0.4cm} s $+= 1$, e $-= 1$
\State \hspace{0.2cm}  \textbf{def sort\_asc($self$):} 
\State \Comment{{\small \textit{Method that modifies the OS}}}
\State \hspace{0.2cm} \hspace{0.2cm}\textbf{for} $i$ \textbf{in} range($self$.get\_size()):
\State \hspace{0.2cm} \hspace{0.3cm} \textbf{for} $j$ \textbf{in} range($i + 1$, $self$.get\_size()):
\State \hspace{0.2cm} \hspace{0.5cm}  \textbf{if} $self$.data$[i]$ $>$ $self$.data$[j]$: \textcolor{red}{\#~mutant: $``<"$}
\State \hspace{0.2cm} \hspace{0.7cm}{\small $self$.data$[i]$, $self$.data$[j] =$ $self$.data$[j]$, $self$.data$[i]$}

\State \hspace{0.2cm} \textbf{def is\_empty($self$)}:
\State \Comment{{\small \textit{Method that inspects the OS}}}
\State \hspace{0.7cm}  \textbf{return} $self$.get\_size() $== 0$
\State \hspace{0.2cm} \textbf{def get\_size($self$):}
\State \hspace{0.2cm} \hspace{0.2cm} size $= 0$ \Comment{{\small \textit{Method that inspects the OS}}}
\State \hspace{0.2cm} \hspace{0.2cm} \textbf{for} element \textbf{in} $self$.data:
\State \hspace{0.2cm} \hspace{0.4cm} \hspace{\algorithmicindent} size $+=$ 1
\State \hspace{0.2cm} \hspace{0.2cm} \textbf{return} size
\State \hspace{0.2cm} \textbf{def get\_first($self$):}
    \State \hspace{0.2cm} \hspace{\algorithmicindent} \textbf{if} $self$.is\_empty(): \Comment{{\small \textit{Method that inspects the OS}}}
    \State \hspace{0.2cm} \hspace{\algorithmicindent} \hspace{\algorithmicindent} \textbf{return} None
    \State \hspace{0.2cm} \hspace{\algorithmicindent} \textbf{else}:
    \State \hspace{0.2cm} \hspace{\algorithmicindent} \hspace{\algorithmicindent} \textbf{return} $self$.data[0]
\State \hspace{0.2cm} \textbf{def get\_last($self$):}
\State \hspace{0.2cm} \hspace{0.2cm} \textbf{if} $self$.is\_empty(): \Comment{{\small \textit{Method that inspects the OS}}}
\State \hspace{0.2cm} \hspace{0.2cm} \hspace{0.2cm} \textbf{return} None
\State \hspace{0.2cm} \hspace{0.2cm} \textbf{else}:
\State \hspace{0.2cm} \hspace{0.2cm} \hspace{0.2cm} \textbf{return} $self$.data$[self$.get\_size() $- 1]$ \textcolor{red}{\#~mutant: $``2"$}
\State \hspace{0.2cm} \textbf{def avg($self$):} 
\State \Comment{{\small Method for OS inspection and manipulation}}
    \State \hspace{0.2cm} \hspace{0.2cm} \textbf{if} $self$.is\_empty(): 
    \State \hspace{0.2cm} \hspace{0.2cm}\hspace{0.2cm} \textbf{return} None
    \State \hspace{0.2cm} \hspace{0.2cm} \textbf{else}:
    \State \hspace{0.2cm} \hspace{0.2cm}\hspace{0.2cm} total = 0
    \State \hspace{0.2cm} \hspace{0.2cm} \hspace{0.2cm}\textbf{for} element \textbf{in} $self$.data:
    \State \hspace{0.2cm} \hspace{0.2cm}\hspace{0.2cm}\hspace{0.2cm} total += element
    \State \hspace{0.2cm} \hspace{0.2cm}\hspace{0.2cm} \textbf{return} total / $self$.get\_size()
\State \hspace{0.2cm} \textbf{def sum($self$):}

    \State \hspace{0.2cm} \hspace{0.2cm}total = 0  \Comment{{\small Method for OS inspection and manipulation}}
    \State \hspace{0.2cm} \hspace{0.2cm}\textbf{for} element \textbf{in} $self$.data:
    \State \hspace{0.2cm} \hspace{0.2cm}\hspace{0.2cm}total += element
    \State \hspace{0.2cm} \hspace{0.2cm}\textbf{return} total

\end{algorithmic}
\end{algorithm}

To illustrate our approach, let us consider a refactoring scenario with a limited test suite. Hasty Sims, a developer, works on a project that involves handling large amounts of data in arrays from different sources. To facilitate the task of performing several operations on these arrays, including calculating the average and sum, reversing the array, sorting, and more, Hasty Sims developed the \texttt{ArrayCalculator} class, \Cref{alg:toyExample}, (hereinafter \texttt{AC}). The \texttt{AC} class contains methods that initialise the object, modify the object's state, inspect the object state, and methods that inspect and perform operations with the object's state data. In  \Cref{alg:toyExample}, the term ``OS" stands for Object State. Below is a detailed description of the methods in \Cref{alg:toyExample}:

\begin{itemize}
    \item Method (``setter") that initialises the object  \texttt{\_init\_(self, data)}
    \item Methods (``setters") that modify the object's state data: \texttt{append\_data(self, new\_data)} which modifies the object's state by adding new data to the existing data in the data instance, \texttt{reverse\_data(self)} which modifies the object's state by reversing the order of the data stored in the data instance, and \texttt{sort\_asc(self)} that modifies the object's state by sorting the data stored in the data instance.

    \item Methods (``getters") that inspect the state data of the object only: \texttt{is\_empty(self)} which inspects the object by checking whether the data instance is empty or not, \texttt{get\_size(self)} that inspects the object by returning the size of the data instance, \texttt{get\_first(self)} and \texttt{get\_last(self)} which inspects the object by returning the first or last element of the data instance, respectively.

    \item Methods (``getters") that inspect and perform operations with the object state data but do not modify the object state: \texttt{get\_avg(self)}, examines the object's state data and calculates the average value, then returns the result, and \texttt{get\_sum(self)}, iterates through the object's state data, adds up all the elements, and returns the total sum value.
\end{itemize}

\begin{algorithm}
\caption{ArrayCalculator test suite}
\label{alg:toyExample_test}
\begin{algorithmic}[1]
\State Require \texttt{unittest} package 
\State \textbf{class} TestArrayCalculator(\texttt{unittest.TestCase}):
\State \hspace{0.2cm} \textbf{def test\_sum$(self$):}
\State \hspace{\algorithmicindent} $self$.array = ArrayCalculator$([2, 3, 1, 4, 5])$
\State \hspace{\algorithmicindent} $self$.\textbf{assertEqual}$(self$.array.sum$(), 15)$
\State \hspace{0.2cm} \textbf{def test\_avg$(self)$:}
\State \hspace{\algorithmicindent} $self$.array = ArrayCalculator$([2, 3, 1, 4, 5])$
\State \hspace{\algorithmicindent} $self$.\textbf{assertEqual}$(self$.array.avg$(), 4.5)$

\end{algorithmic}
\end{algorithm}

\begin{algorithm}
\caption{ArrayCalculator test sequences}
\label{alg:toyExample_test2}
\begin{algorithmic}[1]
\State \hspace{0.2cm} \textbf{def sequence\_one$(self$):}
\State \hspace{\algorithmicindent} $self$.array = ArrayCalculator$([2, 3, 1, 4, 5])$
\State \hspace{\algorithmicindent} $self$.array.append$([6, 7, 8])$
\State \hspace{\algorithmicindent} $self$.array.reverse$()$
\State \hspace{\algorithmicindent} $self$.array.append$([9, 10])$
\State \hspace{\algorithmicindent} $self$.array.sum$()$
\State \hspace{0.2cm} \textbf{def sequence\_two$(self)$:}
\State \hspace{\algorithmicindent} $self$.array = ArrayCalculator$([2, 3, 1, 4, 5])$
\State \hspace{\algorithmicindent} $self$.array.reverse$()$
\State \hspace{\algorithmicindent} $self$.assertEqual$(self$.array.avg$())$
\State \hspace{\algorithmicindent} $self$.array.append$([6, 7, 8])$
\State \hspace{\algorithmicindent} $self$.array.reverse()
\end{algorithmic}
\end{algorithm}

To ensure the functionality of the \texttt{AC} class, Hasty Sims decides to test the \texttt{get\_avg()} and \texttt{get\_sum()} methods, \Cref{alg:toyExample_test}. These calculations are the primary focus of Hasty Sims analysis. Although the other methods, such as \texttt{is\_empty()}, \texttt{get\_size()}, \texttt{get\_first()}, and \texttt{get\_last()}, are also essential, Hasty Sims finds them to be straightforward and not requiring extensive testing.

After Hasty Sims successfully passed the initial tests for the \texttt{AC} class, Sloopy Potter, another developer, made modifications to the class by refactoring the existing methods to optimise their performance. Confident that the functionality of the methods would not be affected, Sloopy Potter did not add any new test cases. After running the old test suite, all the tests passed without errors, leaving Sloopy Potter pleased with their modifications and confident that the class was even more efficient than before. The code written by Sloppy Potter is shown in \Cref{alg:toyExample}. The black code is correct - but assume that Sloppy Potter had injected bugs (marked with the word `mutant' and highlighted in red). Failures triggered by those bugs wouldn't have been spotted by the existing test suite.

Veri Finnick, a diligent developer, wanted to use the \texttt{AC} class. However, she was committed to check that the refactored version of the \texttt{AC} class works correctly. Veri Finnick did not have sufficient time to inspect the changes made or to create additional test cases. Fortunately, Veri Finnick remembered a simple approach for automatically generating amplified regression test oracles using the object state data. 

Veri Finnick meticulously followed the proposed approach, starting with Step 1.1, where she 
employed a tool that randomly generated sequences of method calls. Using this tool and setting its parameters to generating only two test sequences of length five, Veri Finnick generated two test sequences, each comprising four random method calls, as shown in \Cref{alg:toyExample_test2}. Since the original test suite consisting of two unit tests already existed, Veri Finnick to use these tests, too.

\begingroup
\setlength{\tabcolsep}{6pt} 
\renewcommand{\arraystretch}{1} 
\begin{table*}[ht!]
\centering
  \caption{State data of the \texttt{AC} class first version VS State data of the \texttt{AC} class second version}
\label{tbl:toyExampleStates}
\resizebox{\linewidth}{!} {
\begin{tabular}{l|l|l|c|c|c|c|c|c|c|c|c|c|c|c}
\toprule
\multirow{2}{*}{\textbf{test$_{ID}$}}& 
\multirow{2}{*}{\textbf{Input}}& 
\multirow{2}{*}{\textbf{MC$^\star$}} & 
\multicolumn{6}{c|}{\textbf{\texttt{AC} version 1}} & 
\multicolumn{6}{c}{\textbf{\texttt{AC} version 2 (refactored and buggy)}}  \\
 &  &  & get\_last & get\_first & get\_size & is\_empty & sum & average & get\_last & get\_first & get\_size & is\_empty & sum & average \\
\midrule
\multirow{2}{*}{test\_sum} & \multirow{2}{*}{[2, 3, 1, 4, 5]}	&	
CTOR	&	-	&	-	& -	&	-	&	-	&	-	&	-	&	-	&	-	&	-	&	-	&	-	\\
& 	&	
sum	    &	5	&	2	& 5	&	FALSE	&	15	&	3	&	5	&	2	&	5	&	FALSE	&	15	&	3	\\
\midrule
\multirow{2}{*}{test\_avg}  & \multirow{2}{*}{[2, 3, 1, 4, 5]}	&	CTOR	&	-	&	-	& -	&	-	&	-	&	-	&	-	&	-	&	-	&	-	&	-	&	-	\\
& &	average	&	5	&	2	& 5	&	FALSE	&	15	&	3	&	5	&	2	&	5	&	FALSE	&	15	&	3	\\
\midrule
\multirow{5}{*}{sequence\_one} & \multirow{5}{*}{[2, 3, 1, 4, 5]}	&	CTOR	&	-	&	-	& -	&	-	&	-	&	-	&	-	&	-	&	-	&	-	&	-	&	-	\\	
& &	append	& \textbf{8}	&	2	& 8	&	FALSE	&	36	&	4.5	&	\textbf{7}	&	2	&	8	&	FALSE	&	36	&	4.5	\\
& &	reverse	&	\textbf{2}	&	\textbf{5}	& 5	&	FALSE	&	15	&	3	&	\textbf{4}	&	\textbf{2}	&	5	&	FALSE	&	15	&	3	\\
& &	append	&	\textbf{10}	&	\textbf{5}	& 10&	FALSE	&	55	&	5.5	&	\textbf{9}	&	\textbf{2}	&	10	&	FALSE	&	55	&	5.5	\\
& &	sum	&	\textbf{10}	&	\textbf{5}	& 10&	FALSE	&	55	&	5.5	&	\textbf{9}	&	\textbf{2}	&	10	&	FALSE	&	55	&	5.5	\\

\midrule

\multirow{5}{*}{sequence\_two} & \multirow{5}{*}{[2, 3, 1, 4, 5]} &	CTOR	&	-	&	-	& -	&	-	&	-	&	-	&	-	&	-	&	-	&	-	&	-	&	-	\\
& &	reverse	&	\textbf{2}	&	\textbf{5}	& 5	&	FALSE	&	15	&	3	&	\textbf{4}	&	\textbf{2}	&	5	&	FALSE	&	15	&	3	\\
& &	avg	&	\textbf{2}	&	\textbf{5}	& 5	&	FALSE	&	15	&	3	&	\textbf{4}	&	\textbf{2}	&	5	&	FALSE	&	15	&	3	\\
& &	append	&	\textbf{8}   &	\textbf{5}	& 8	&	FALSE	&	36	&	4.5	&	\textbf{7}	&	\textbf{2}	&	8	&	FALSE	&	36	&	4.5	\\
& &	reverse	&	\textbf{5}	&	\textbf{8}   & 8	&	FALSE	&	36	&	4.5	&	\textbf{7}	&	\textbf{2}	&	8	&	FALSE	&	36	&	4.5	\\

\bottomrule
\multicolumn{5}{l}{$^\star$Method call during test execution}
\end{tabular}}
\end{table*}%
\endgroup

For Step 1.2, Veri Finnick selected all the methods that have a return value, to be monitored by the Test Adapter. This means that these methods will be executed after each sequence of both test suites (\Cref{alg:toyExample_test} and \Cref{alg:toyExample_test2}). Then, following step 1.3, Veri Finnick executed both test suites on the original code without any test failing and obtained the output from the Test Adapter, which was a CSV file containing the state data. Veri Finnick continued with the approach by executing steps 2.1 and 2.2 with the refactored (buggy) version of the class. She could run both test suites without any test failures and obtained the CSV file containing the state data of the second version. The object state data captured by Test Adapter during test execution is presented in \Cref{tbl:toyExampleStates}. The Test\_ID column provides the identification of the executed test, while the input column displays information on the initialisation value of the object. The MC column provides information on the called method, and the six methods used to monitor the state are grouped according to version 1 (original) and version 2 (refactored and buggy) of the \texttt{AC} class.

After comparing the two versions, Veri Finnick quickly realized that the behaviour of the second version was different from the first one. To identify the root cause of this difference, Veri checked the adapter methods with different behaviour. During this process, she found that the \texttt{get\_last} method in the second version returned the wrong position, indicating the presence of a bug. Veri then investigated the \texttt{get\_first} method but could not find any issues with it. Veri also noticed that in sequence\_one, after the \texttt{reverse} method was called, the behaviour of the entire sequence for the \texttt{get\_first} method was different. This made her suspicious that the \texttt{reverse} method might be the cause of this unexpected behaviour. She investigated further and discovered a second bug in the \texttt{reverse} method, which was causing the \texttt{get\_first} method to provide the wrong state data. After fixing the bugs, Veri reran the fixed version and compared it with the first version once again. To her satisfaction, everything was identical to the behaviour of the first version, indicating that the bugs were successfully resolved.

We aim to highlight three main points with this toy example. Firstly, if the initial test suite had at least one test for each method that has a return value, it is likely that the bug in the \texttt{get\_last} method could have been found through the test suite. However, even with such a test suite, the second bug that was found with the help of the Test Adapter may not have been caught. This is because the bug was not related to the return values of the method but to its side effects on the object's state. Secondly, it is important to note that there may be cases where neither the Test Adapter nor the test suite is able to find a bug. For instance, in our toy example, the bug injected in the \texttt{sort\_asc} method was not caught by either the Test Adapter or the test suite. Nevertheless, this bug could have been easily found by Test Adapter because of the potential side effects it could have on the object's state data if called in a certain sequence of tests. Thirdly, while it is true that a more comprehensive and robust test suite can render our approach obsolete, it is important to note that even with a strong test suite, the information about the state data can still be valuable for debugging and fault localisation, as was demonstrated by Veri Finnick in our toy example. By examining the object state data captured during the test execution for the SUT's faulty and corrected versions, Veri Finnick could pinpoint the bug's exact location in the \texttt{get\_last} method.

It is important to note that our approach is not meant to replace existing test suites. Instead, it complements them by providing additional information for more effective regression testing. By comparing the state information captured during the test execution of two different versions of the SUT, our approach can amplify the existing test suite's ability to identify faults and can help facilitate the process of debugging.

\subsection{Evaluation} 
\label{subsubsec:SUT}
We addressed our research question by implementing our approach on a custom implementation of the Collection library in Java as the SUT. We conducted a mutation testing analysis to evaluate the effectiveness of our approach. The Collection library comprises 22 Java files,  and approximately 1400 lines of code within the \texttt{experimental.util} package. It features implementations of several common collection algorithms, including \texttt{Stack}, \texttt{ArrayList}, \texttt{LinkedList}, \texttt{HashMap}, \texttt{TreeMap}, \texttt{HashSet}, and \texttt{TreeSet}. The library's classes were developed using object-oriented principles, with defined interfaces and abstract implementations to enable code reuse and inheritance. However, our experiment did not include static classes that provide common functionality, as they do not maintain state data. The full set of data generated during our experiments as well as all scripts, can be found in our GitHub repo\footnote{https://github.com/software-competence-center-hagenberg/test-amplification-SEAA2023}.

\section{Results and discussion}
\label{sec:results_discussion}

\subsection{RQ$_1$: How can object state data be used to derive test oracles during test execution?}
\label{subsec:RQ1}
This research question examines the practicality of using object state data to create test oracles during test execution. The aim is to investigate whether it is possible to monitor the state of objects during the execution of the SUT and capture and store their state data at different stages of the test execution. In the following sub-sections, we provide a detailed description of the implementation of our approach, with a special emphasis on the generation of the Test Adapter.


\textbf{\textit{Step 1.1 - Produce initial test:}} We employed Randoop, a well-known random unit test generator for Java, to generate an initial test suite for our experiments. Randoop randomly selects a method call to apply and uses arguments from previously constructed sequences to generate method sequences. 
In our experiments, we used two key parameters, namely the test limit and the random seed, to control the size and variability of the generated test suites. 
To ensure that the generated test suite with the higher number of test sequences contains all the test cases of a test suite with lower nuber of test sequences, we decided to split the creation of the initial test classes into two steps. In the first step we utilised Randoop to generate a test class with a maximum of 1024 test sequences for every generated Test Adapter and repeated it for 10 different seed numbers, which resulted in a total number of 70 generated test classes.
Afterwards we split every generated test classes with the help of a utility function into 9 separate test classes containing the specified number of test sequences (2, 4, 8, 16, 32, 64, 128, 256, 512). This resulted in the total number of 630 test classes, that is, 70 per above mentioned test sequence number limit.
For more information about the generation of the test classes and test sequence limits, please refer to the detailed information available at our repo.


We used a batch script and Randoop for every driver class, which represents the Test Adapter that allows us to capture the internal state data, and seed with a configured test number creation limit of 1024. We utilised Randoop options that allowed us to specify code snippets to be executed before and after every test case execution. The code snippet executed before the test execution created a unique test identifier and provided it as a system property to be easily accessible by the generated tests. 

\textbf{\textit{Step 1.2 - State data acquisition:}} As mentioned in previous sections, the crucial aspect of our approach is recording the SUT's state data after each method call during the test execution. We extracted information from public methods of the classes that do not modify the state data of the class and used this information to configure the interface for test generation. For example, in the \texttt{Stack class}, the \texttt{peek} method does not modify the state data of the stack instance and can therefore be used as a source of information about its state data. On the other hand, the \texttt{pop} method removes the last object from the instance and may not be suitable for retrieving the current state. Although it may be possible to automatically identify state data-preserving methods through static and dynamic code analysis, this was not the focus of our research. We manually inspected the code to extract what methods will be used by the Test Adapter. To record the SUT's state data, the Test Adapter forwards calls to the SUT and records its state data. We implemented a solution that automatically generates this layer of indirection with just a few configuration parameters.



The configuration file is a CSV file that contains the methods that Test Adapter will use. Each line represents a target class of the SUT and includes the fully qualified name of the class, a listing of optional public methods (only used during development), and a space-separated list of public methods that represent (part of) the state data of the SUT. The Test Adapter, represented by the `AdapterClassGenerator', uses this file as input and creates seven Adapter classes, each named with the appended term \texttt{TestDriver} (\textit{e.g.,} \texttt{StackTestDriver} for class Stack). These classes provide additional functionality to record and compare the state data for every executed test. The Adapter uses Java reflection capabilities to retrieve and filter the available inherited public interface and feeds a template engine with corresponding information to create the test drivers. The output location of the generated test driver classes can be specified using an optional command-line program argument.

\textbf{\textit{Step 1.3 - Test execution:}}  To execute all the tests, we used a batch script for every driver class. Using a batch script, the execution of the generated test cases was automated, which helped to save time and effort. The script allowed for the execution of a large number of test cases without the need for manual intervention, reducing the chance of errors and ensuring consistent test results. 


\textbf{\textit{Step 2.1 - re-execution of test suite:}} Assuming that a new version of the SUT has been generated, we would need to rerun all the tests from Phase I to ensure that they are compatible with the updated version. We would then save the state of each test to compare it with the state recorded during Phase I. 

\textbf{\textit{Step 2.2 - Amplified regression test oracle:}} The Adapter class created during step 1.2 not only forwards the method calls to the class under test but also records the state data of the class after the method call is executed. This is achieved by calling the method \texttt{writeInternalState} with the name of the executed method and its arguments. 


To compare the current test run's state data entries with an existing file or save the state as the new expected state file, the generated Adapter classes provide the static method \texttt{matchInternalStateSnapshot}. This method can be called after the execution of a test. Furthermore, the comparison can be enabled or disabled via a system property required for our research experiment.


\begin{tcolorbox}[top=4pt, left=4pt, right=4pt]
We have addressed RQ$_1$ by outlining the steps required to use object state data to create amplified regression test oracles during test execution. This involves generating Test Adapter, executing tests generated by Randoop, and recording the state data of the SUT using Test Adapter. We have also shown how to compare the recorded state data with the expected results to generate test oracles. This step-by-step process highlights the practical feasibility of using object state data for regression test oracle derivation.
\end{tcolorbox}

\subsection{RQ$_{2}$: What is the effectiveness of the amplified regression test oracles in detecting unintended behaviour in the SUT?} 
\label{subsec:RQ2}

We conducted an experiment to measure the benefits of using amplified regression test oracles for detecting unintended behaviour in the SUT. We used the pitest\footnote{https://pitest.org/} tool, a Java mutation testing tool, to execute the experiment. Pitest seeded mutations in the SUT and executed only the necessary tests to cover those mutations, generating a log file and an HTML report. The report provided coverage, mutation score, and test strength at the package and test class level. We applied mutation testing twice for all tests grouped by seed numbers (10) and test limit numbers (9), once without following our approach and once with our approach.
This resulted in $180$ mutation test runs for our experimental setup, which took about 18.5 hours to execute and involved about 1.3 million test sequences.

We automated mutation tests using a batch script that looped over the used seeds and limits. Pitest supported filtering test classes by name using wildcards to consider all generated test classes for a given seed and test limit in a mutation test run. Reports from mutation test runs were written to a configured report directory named by concatenating the current seed and limit values. The mutation batch script executed successfully, creating $90$ report directories with reports for runs using our approach and without. This experiment provided us with insights into the potential benefits of using amplified regression test oracles to detect unintended behaviour in the SUT.

\begin{figure}[tp!]
	\centering
 	\includegraphics[trim= 6mm 1mm 15mm 11mm, width=\linewidth]{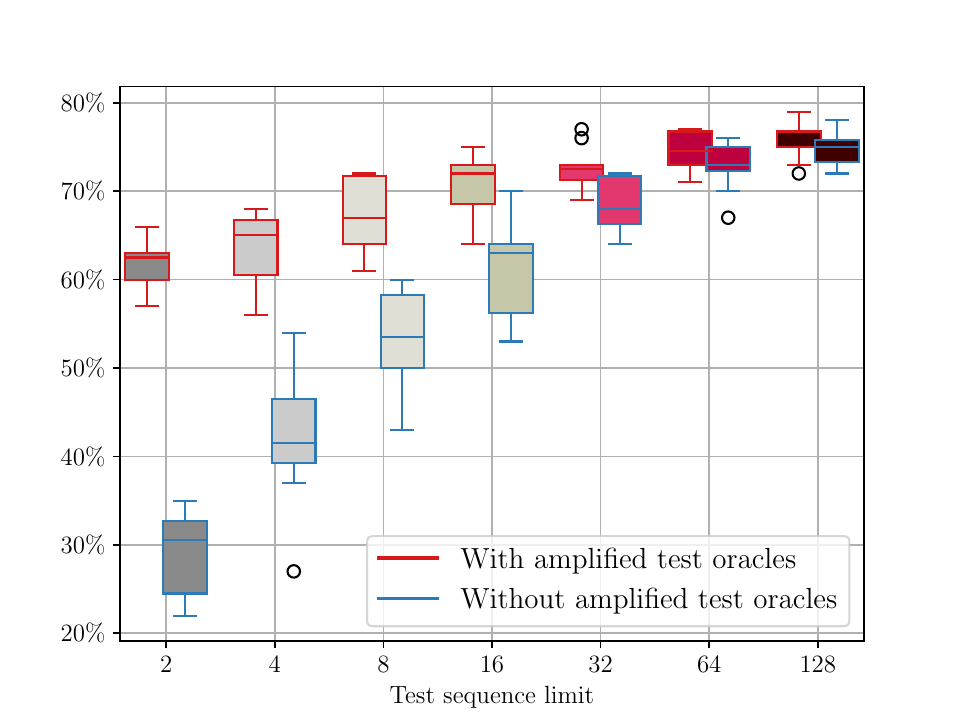}
	\caption{Test strength per test limit}
\label{fig_diagram}
\end{figure}

Our initial results showed that our approach had the most potential when dealing with a limited number of test sequences. \Cref{fig_diagram} provides a clear visual representation of the relationship between test strength and the number of executed tests, grouped by specific test limit number. It shows that our approach significantly enhances test strength by incorporating state data information of the SUT, thereby improving the ability of tests to detect changes.

It is important to note that as the number of tests increases, there may be a point where the Test Adapter's capabilities are reached (converging results in the figure), meaning that the test suite will have the same bug-finding capabilities as the Test Adapter. However, even when the test suite reaches its maximum capabilities, incorporating the state data of the SUT can still provide helpful information for debugging and localising bugs. The state data provides a detailed snapshot of the behaviour of the SUT during the test execution, which can help narrow down the cause of the problem. By analysing the differences between the state data, developers/testers can quickly identify where the code has changed and focus their debugging efforts on those areas.

For a more detailed discussion of the evaluation results, please refer to \Cref{tbl:MutationResultDetails}. 
Since our evaluation results show that our approach does not improve the bug-finding capabilities for test classes containing more than 128 test sequences, we have chosen to discuss the results for the two lower test sequence number limits. Mutation testing results are presented in groups based on the number of test sequences, namely 2, 16, and 128. For each group, the appropriate column contains the average values calculated once without and once with our approach. For lower test sequence limits, there is no much difference in the execution time but it increases strongly for higher test sequence number limits. This was attributed to the fact that the current implementation persists the state to files and reads it from files for comparison, causing slow file operations. 
Upon examining the average number of tests executed during the mutation test run, one may notice that the number decreases when object state data comparison is enabled. This is because Pitest stops executing other tests as soon as the mutant covered by these tests has been killed.
The increase in the number of killed mutants corresponds to the higher percentage of the test strength and shows an improvement at a test sequence number limit of 2 from 28.9\% to 61.9\% on average, which is equivalent to an increase of 114\%. But although the results show that the lead of our approach vanish for a higher number of existing test sequences, we think that is mainly caused by the simplistic approach we have chosen to track and record the object state. Therefore we would like to encourage researchers to develop a more powerful way to track and use the system's internal state for regression testing.

\begin{table}[ht]
\caption{Mutation Testing Results by Test Sequence Limit}
\label{tbl:MutationResultDetails}
\centering
\resizebox{\linewidth}{!} {
\begin{tabular}{l|cc|cc|cc}
\toprule
\multirow{2}{*}{{\scriptsize \diagbox{\textbf{Metrics}}{\textbf{TSL$^\star$}}}} & \multicolumn{2}{c|}{\textbf{2}} & \multicolumn{2}{c|}{\textbf{16}} & \multicolumn{2}{c}{\textbf{128}}\\
& WATO$^\dagger$ & W/O ATO$^\pm$ & WATO$^\dagger$ & W/O ATO$^\pm$ & WATO$^\dagger$ & W/O ATO$^\pm$ \\
\midrule
Execution time $(s)$ & $38$ & $45$ & $171$ & $199$ & $206$ & $445$  \\
\midrule
Executed tests & $166$ & $136$ & $1668$ & $1169$ & $9572$ & $8381$  \\
\midrule
Killed mutants & $33$ & $71$ & $162$ & $187$ & $232$ & $234$  \\
\midrule
Test strength $(\%)$ & $28.9$ & $61.9$ & $61.3$ & $70.7$ & $74.7$ & $75.5$  \\
\bottomrule
\multicolumn{7}{l}{\textbf{$^\star$}Test Sequence Limit} \\
\multicolumn{7}{l}{\textbf{$^\dagger$}With Amplified Test Oracles, \textbf{$^\pm$}Without Amplified Test Oracles } \\
\end{tabular}}
\end{table}


\begin{tcolorbox}[top=4pt, left=4pt, right=4pt]
Regarding RQ$_2$, our results showed that using our approach, we could more than double the number of killed mutants at best, increasing average test strength from 28.9\% to up to  61.9\% on. Incorporating the object state data of the SUT significantly increased test strength, improving the tests' ability to detect changes. Furthermore, even when the test suite reached its maximum capabilities, using the object state data of the SUT could still provide helpful information for debugging and localising bugs.
\end{tcolorbox}

\subsection{Threads of Valdity}
\label{validity}
\subsubsection{Internal validity} Specific tools and configurations for the test code generation and assessment of the generated tests have been used. 
We tried to at least mitigate the problem of randomness when generating test classes by using different values for the seed. Application of more sophisticated test case generator tools, or using other configurations, might provide different results for the test strength. Furthermore, we used pitest to assess and compare the change detection capability only. Errors in the tools might bias our results, although the measured difference between the different test series should be valid because the same toolchain has been used for all measurements.

\subsubsection{External validity}
Our approach has been implemented only for one single programming language and SUT. 
Also, our approach was applied to one simple experimental software system. Studies on more systems are required to determine the degree to which results may be generalised. Nonetheless, our results indicate that if a system allows the observation of the state it should improve the change detection capability.

\section{Related work}
\label{sec:related_work}
A review of methods for automatically generating test oracles can be found in \cite{10.1145/3266237.3266273} and \cite{6963470}. Several approaches have been proposed for test amplification in regression testing, such as Orstra \cite{10.1007/11785477_23} and DSpot \cite{paper}, which aim to augment JUnit test suites with regression oracle checking. \citeauthor{abdi2019test} \cite{abdi2019test} propose Small-Amp, an amplification approach for the Pharo Smalltalk ecosystem that builds on the DSpot framework. \citeauthor{https://doi.org/10.1002/smr.2490} \cite{https://doi.org/10.1002/smr.2490} provide similar tools for Python. In contrast to existing amplification approaches that target programs implemented in specific languages such as Java \cite{paper}, Pharo Smalltalk \cite{abdi2019test}, and Python \cite{https://doi.org/10.1002/smr.2490}, our approach stands out from existing amplification approaches by utilizing object state data to create regression test oracles during test execution. By inspecting the state data, we can generate additional information for amplified regression test oracles, which can significantly improve the effectiveness of the testing process. This feature sets our approach apart as it allows us to amplify the existing test suite for any program, regardless of the programming language.

\section{Conclusion and future work}
\label{sec:conclusion_future_work}

In this paper, we proposed an approach to amplify automatically generated test cases by monitoring the object state data of the SUT during test execution and comparing it with the previous version of the SUT to detect changes. Our research aimed to answer two questions: whether object state data can be used to derive test oracles during test execution and whether amplified regression test oracles are effective in detecting unintended behaviour in the SUT. Our experiments on Collection Classes in Java showed that monitoring state data improved test accuracy and provided helpful information for debugging and locating bugs.

We recognise that the automatic extraction of the interface representing the state data of a system is a challenge that needs to be addressed in future research. Moreover, we intend to enhance our test oracle amplification approach by leveraging novel learning technologies. Instead of executing specific, pre-recorded state checks for regression testing, we aim to develop a relationship model that enables us to derive generalised rules. These rules can then be automatically injected and checked during test execution using the SUT's current internal state. This could offer a potential solution to the maintainability problem of automatically generated or modified test cases.

\section*{Acknowledgement}
The research reported in this paper has been partly funded by BMK, BMAW, and the State of Upper Austria in the frame of the SCCH competence center INTEGRATE [(FFG grant no. 892418)] part of the FFG COMET Competence Centers for Excellent Technologies Programme, as well as by the European Regional Development Fund, and grant PRG1226 of the Estonian Research Council.

\printbibliography

\end{document}